\documentclass[aps,twocolumn,showpacs,floats]{revtex4}
\usepackage{graphicx}
\usepackage{dcolumn}
\usepackage{bm}
\usepackage{color}
\usepackage{amsmath}
\usepackage{amssymb}

\begin{document}

\title{Anderson localization of matter waves in quantum-chaos theory}

\author{E. Fratini and S. Pilati$^{1}$}
\affiliation{$^{1}$The Abdus Salam International Centre for Theoretical Physics, 34151 Trieste, Italy}

\begin{abstract}
We study the Anderson localization of atomic gases exposed to three-dimensional optical speckles by analyzing the statistics of the energy-level spacings.
This method allows us to consider realistic models of the speckle patterns, taking into account the strongly anisotropic correlations which are realized in concrete experimental configurations.
We first compute the mobility edge $E_c$ of a speckle pattern created using a single laser beam. We find that $E_c$ drifts when we vary the anisotropy of the speckle grains, going from higher values when the speckles are squeezed along the beam propagation axis to lower values when they are elongated.
We also consider the case where two speckle patterns are superimposed, forming interference fringes, and we find that $E_c$ is increased compared to the case of idealized isotropic disorder.
We discuss the important implications of our findings for cold-atom experiments.
\end{abstract}

\maketitle


Anderson localization is the complete suppression of wave diffusion due to destructive interferences induced by sufficiently strong disorder~\cite{wiersma}. 
It was first discussed by Anderson in 1958~\cite{anderson} and has been observed (only much later) in various physical systems, including light waves~\cite{maret1,maret2,roghini,segev}, sound waves~\cite{tiggelen}, and microwaves~\cite{genack}, and also in experiments performed with ultracold gases, first implementing an effective Anderson model~\cite{garreau}, and then observing the localization of matter waves in one dimension~\cite{aspect,inguscio1} and in three dimensions~\cite{aspect2,demarco1}.
Recently, transverse Anderson localization has been realized in randomized optical fibers~\cite{karbasi1}, paving the way to potential applications in biological and medical imaging~\cite{karbasi2}.\\
The key quantity which characterizes Anderson localization in three-dimensional quantum systems is the mobility edge $E_c$, which is the energy threshold that separates the localized states (with energy $E < E_c$) from the delocalized ones (with energy $E > E_c$)~\cite{ramakrishnan}.
Many accurate theoretical predictions for the value of $E_c$ exist, but most of them regard simplified toy models defined on a discrete lattice~\cite{kramer,romer}. These lattice models do not describe the spatial correlations, and their possible anisotropy, of the disorder present in the physical systems where Anderson localization has been observed. 
In fact, these features are expected to have a profound impact on the Anderson transition. For example, it is known that due to finite spatial correlations an effective mobility edge exists also in low-dimensional systems~\cite{krokhin,palencia2007,palencia1,lugan,kenneth}, while for uncorrelated disorder all states would be localized~\cite{ramakrishnan}. According to recent results~\cite{hilke}, in continuous-space systems localization does not occur if the disorder correlation length vanishes, even for strong disorder. It is also known that the structure of the spatial correlations changes drastically the localization length and the transport properties~\cite{piraud1,piraud3}.\\
The experiments performed with ultracold atoms are emerging as the ideal experimental setup to study Anderson localization~\cite{inguscio2,lewenstein,shapiro}.
Different from other condensed-matter systems, atomic gases are not affected by absorption effects and permit us to suppress the interactions. 
Furthermore, by shining coherent light through diffusive surfaces, experimentalists are able to create three-dimensional disordered profiles (typically referred to as optical speckle patterns) with tunable intensity and to manipulate the structure of their spatial correlations.\\

\begin{figure}
\begin{center}
\includegraphics[width=1.0\columnwidth]{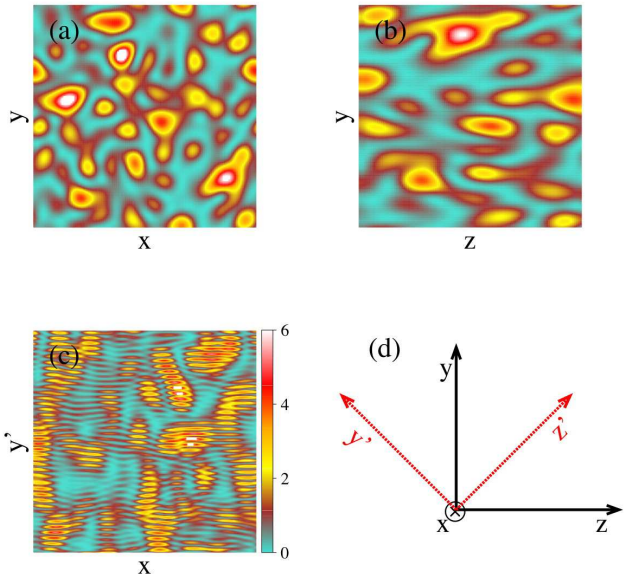}
\caption{(color online) 
(a) Intensity profile of a speckle pattern measured on a plane orthogonal to the beam propagation axis $z$. 
(b) Elongated speckle pattern with anisotropy $\sigma_z/\sigma=6$, measured along a plane containing the beam axis $z$.
(c) Profile resulting from two orthogonally crossed speckle patterns (see text) measured on a plane containing the second principal axis.
(d) Representation of the two speckle-patterns configuration, indicating the propagation directions of the first beam ($z$) and the second beam ($y$). The red arrows indicate the 1$^{\textrm st}$ principal axis ($z'$) and the 2$^{\textrm nd}$ p. a. ($y'$). The $x$-axis enters the sheet plane.
}
\label{fig1}
\end{center}
\end{figure}

%
In this Rapid Communication, we investigate the Anderson localization of noninteracting atomic gases moving in three-dimensional optical speckles. We determine the single-particle energy spectrum using large-scale diagonalization algorithms. Then, by performing a statistical analysis of the spacings between consecutive energy levels, we locate the mobility edge.
The study of the level-spacing statistics lies at the heart of random-matrix and quantum-chaos theories. It has permitted us to interpret the complex spectra of large nuclei, atoms, and molecular systems~\cite{mehta,haake}. More recently, it has been employed in the analysis of the Google Matrix~\cite{shepelyansky1,shepelyansky2}.
Quantum-chaos theory provides a universal basis-independent criterion for the localization transition. One has to identify two kinds of level-spacing distributions, namely, the Wigner-Dyson distribution characteristic of ergodic chaotic systems, and the Poisson distribution characteristic of localized quantum systems.
This method has allowed researchers to locate the localization transition in noninteracting three-dimensional lattice models (both isotropic and anisotropic)~\cite{shore,hofstetter,schreiber,schweitzer}, and, more recently, also in interacting one-dimensional spin systems~\cite{huse,cuevas,alet,scardicchio}.
In the present study this criterion is used to investigate the Anderson localization of matter waves, setting the basis for future investigations of many-body localization in interacting three-dimensional Fermi gases.\\

First, we consider the experimental configuration with a single speckle pattern created by shining a laser through a diffusive plate. In this case the spatial correlations of the disorder are intrinsically strongly anisotropic, with cylindrical symmetry around the beam propagation axis.
We find that, when the speckles are elongated along the axis, which is the typical experimental situation, the mobility edge is only moderately reduced compared to the idealized models of disorder with spherically symmetric correlations. This unexpected result indicates that the experimental setup with a single speckle pattern is quite suitable to investigate Anderson localization, despite the strong disorder anisotropy.
We also consider the case where two orthogonal speckle patterns are coherently superimposed. This setup, which was originally implemented to avoid the large axial correlation length of the single-pattern configuration, generates an intricate correlation structure, with rapid oscillations of the external field due to interference fringes (see Fig.~\ref{fig1})~\cite{semeghini}.
In this case we find that the mobility edge is higher than for isotropic disorder, and is similarly to the case of a single speckle pattern with axially squeezed speckle grains. This means that the two-pattern configuration provides experimentalists a handle to shift upwards the position of the mobility edge.\\

The first step in the determination of $E_c$ is to compute the spectrum of the single-particle Hamiltonian $\hat{H} = -\frac{\hbar^2}{2m}\Delta + V({\bf r})$, where $\hbar$ is the reduced Planck's constant, $m$ is the atom's mass, and $V({\bf r})$ is the disordered potential experienced by the atoms exposed to optical speckle patterns. We consider a large box with periodic boundary conditions, which has a cubic shape (of size $L$) and parallelepiped shape, for isotropic and anisotropic speckles, respectively.
We tackle this challenging computational task by representing $\hat{H}$ in momentum space, truncating the Fourier expansion at a large wave vector, carefully analyzing that the basis-truncation error is smaller than the final statistical uncertainty. To compute the eigenvalues we employ advanced numerical libraries for high-performance computers with shared-memory architectures~\cite{plasma}. 
 For more details on the Hamiltonian representation and on the numerical diagonalization procedure, see the Supplemental Material~\cite{SM}.\\
If the speckle field is blue detuned with respect to the atomic transitions, it generates a repulsive potential with an exponential probability distribution of the local intensity, which reads $P_{\textrm{bd}}(V) = \exp\left(-V/V_0\right)/V_0$, if the intensity is $V>0$ and $P(V)=0$ otherwise. Thus, the potential has the lower bound $V({\bf r})=0$, while it is unbounded from above. The disorder strength is determined by the energy scale $V_0$, which is equal to the spatial average of the potential, $V_0=\left<V({\bf r})\right>$ and also to its standard deviation, so that $V_0^2=\left<V({\bf r})^2\right>-\left<V({\bf r})\right>^2$. For sufficiently large systems the disorder is self-averaging, and the spatial average coincides with the average over disorder realizations.
Another fundamental property which characterizes the speckle pattern is the two-point spatial correlation function $\Gamma({\bf r}) = \left< V({\bf r}'+{\bf r}) V({\bf r}')\right>/V_0^2-1$. After averaging over the position of the first point ${\bf r'}$, it depends on the relative (vector) distance ${\bf r}$.\\
%
%
\begin{figure}
\begin{center}
\includegraphics[width=1.0\columnwidth]{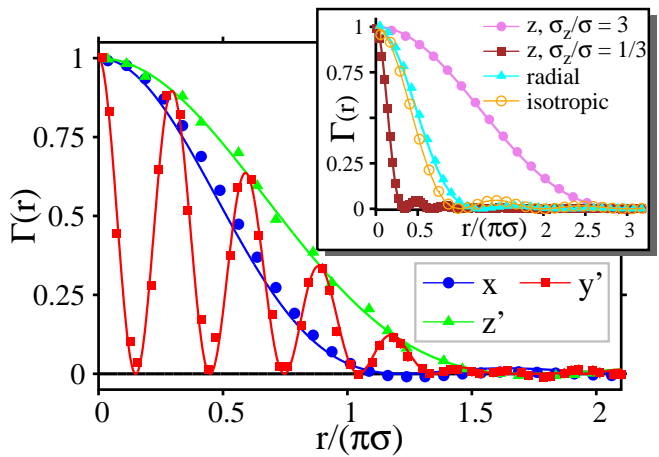}
\caption{(color online) Two-point spatial correlation functions of the disorder. The continuous curves represent the analytical formulas~\cite{goodman} and the symbols represent the correlation measured on the speckle patterns generated numerically.
The inset shows the correlation function of a single speckle-pattern along the beam axis $z$ for elongated speckle grains ($\sigma_z/\sigma=3$) and squeezed speckle grains ($\sigma_z/\sigma=1/3$), radial correlation, and isotropic correlation of idealized spherically-symmetric speckle patterns.
The main panel shows the correlation of crossed speckle patterns along the first ($z'$) and the second principal axis ($y'$) and along the orthogonal axis $x$.}
\label{fig2}
\end{center}
\end{figure}
%
%
In order to make a direct comparison with a previous theoretical study based on transfer-matrix theory~\cite{orso}, we first consider an idealized isotropic model of the speckle pattern with a spherically symmetric correlation function that reads $\Gamma^{\textrm{iso}}(r) = \left[\sin(r/\sigma)/(r/\sigma)\right]^2$ (see inset in Fig.~\ref{fig2}). The parameter $\sigma$ fixes the length scale of the spatial correlations and therefore the typical grain size~\cite{SM}. An efficient numerical algorithm to generate isotropic speckle patterns is described in details in Refs.~\cite{huntley,modugnomichele}.
%
%
%
\begin{figure}
\begin{center}
\includegraphics[width=1.0\columnwidth]{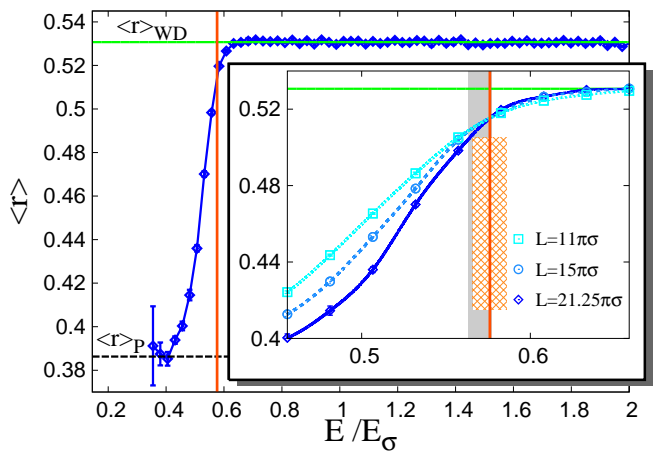}
\caption{(color online). The main panel shows the ensemble-averaged adjacent-gap ratio $\left<r\right>$ as a function of the energy $E/E_\sigma$ for an isotropic speckle pattern of intensity $V_0=E_\sigma$, where $E_\sigma$ is the correlation energy. The horizontal green line is the result for the Wigner-Dyson distribution $\left<r\right>_{\textrm{WD}}$, and the dashed black line the one for the Poisson distribution  $\left<r\right>_{\textrm{P}}$.
The inset gives comparison between different system sizes. The vertical orange line indicates the position of the mobility edge $E_c$ (the hatched rectangle represents the error-bar). The gray bar represents the value of $E_c$ predicted in Ref.~\cite{orso} using transfer-matrix theory.
}
\label{fig3}
\end{center}
\end{figure}
%
We determine the energy spectrum of a large number of realizations of the speckle pattern~\cite{SM}. In the high-energy regime, the energy levels $E_{n}$ (listed in ascending order) fluctuate, avoiding each other, signaling the level repulsion typical of delocalized chaotic systems. The distribution of the level spacings $\delta_n=E_{n+1}-E_{n}$ should correspond to the statistics of random-matrix theory (in particular, to the Gaussian orthogonal ensemble), namely, the Wigner-Dyson distribution. Instead, in the low-energy regime the energy levels easily approach each other like independent random variables. This is a consequence of the localized character of the corresponding wave functions. In this regime the level spacings follow a Poisson distribution. 
In order to identify the two statistical distributions and determine the energy threshold $E_c$ which separates them, we compute the ratio of consecutive level spacings: $r = \min \left\{ \delta_n, \delta_{n-1} \right\}/\max \left\{ \delta_n, \delta_{n-1} \right\}$. The average over disorder realizations is known to be $\left< r \right>_{WD} \simeq 0.5307$ for the Wigner-Dyson distribution, and $\left< r \right>_{P} \simeq 0.38629$ for the Poisson distribution~\cite{roux}. This statistical parameter was first introduced in Ref.~\cite{huse} in the context of many-body localization.
In Fig.~\ref{fig3} we show the data corresponding to the disorder strength $V_0 = E_\sigma$, where $E_\sigma = \hbar^2/m\sigma^2$ is the correlation energy. 
We find that the ensemble average $\left< r \right>$ changes rapidly from $\left< r \right>_{P}$ to $\left<r\right>_{WD}$ as the energy increases.
While in an infinite system one would have a sudden transition between the two statistics (with a third distribution exactly at  $E_c$~\cite{kravtsov1}), in a finite system we have a rapid but continuous crossover. 
For energies $E<E_C$, the data drift towards $\left< r \right>_{P}$ as the system size $L$ increases since the localized wave functions are independent only for $L\rightarrow \infty$, while they drift in the opposite direction for $E>E_c$. The crossing of the curves corresponding to different system sizes indicates the critical energy (see inset of figure~\ref{fig3}).  To pinpoint $E_c$ we fit the data close to the transition with the scaling Ansatz $\left< r \right> = g\left[ \left(E-E_c\right)L^{1/\nu}\right]$~\cite{50years}, where $\nu$ is the critical exponent of the correlation length and $g[x]$ is the scaling function (universal up to a rescaling of the argument) which we Taylor expand up to second order. 
For the case of Fig.~\ref{fig3}, from the best-fit analysis we obtain $E_c = 0.576(10)E_\sigma$, in quantitative agreement with the result of transfer-matrix theory from Ref.~\cite{orso}: $E_c = 0.570(7)E_\sigma$. For the critical exponent we obtain $\nu = 1.6(2)$, which is consistent with the prediction for the Anderson model: $\nu = 1.571(8)$~\cite{slevin}.
It is worth mentioning that in the energy regime $E\sim V_0$ classical particles would be completely delocalized since the energy threshold $\epsilon_p$ for classical percolation in three-dimensional speckle patterns is extremely small, namely, $\epsilon_p\sim 10^{-4} V_0$~\cite{pilati1}.
We consider also a red-detuned speckle field. Its distribution of intensities $P_{\textrm{rd}}(V)$ is the opposite of what corresponds to blue-detuned speckles, that is, $P_{\textrm{rd}}(V) = P_{\textrm{bd}}(-V)$. The corresponding average value is $\left<V(\bf{r})\right>=-V_0$. 
At the disorder strength $V_0=E_\sigma$, we obtain the mobility edge $E_c = -0.81(4)E_\sigma$, which (marginally) agrees with the result of transfer-matrix theory: $E_c = -0.863(6)E_\sigma$~\cite{orso}.
It is worth noticing that for blue-detuned speckles the mobility edge is well below the average intensity of the potential, while for red-detuned speckles it is instead above it. This strong asymmetry was already found in Ref.~\cite{orso} using transfer-matrix theory, but it was not captured by previous approximate calculations based on the self-consistent theory of localization. This means that predicting the position of the mobility edge requires quantitatively accurate methods.\\

We now turn the discussion to concrete experimental configurations.
We first consider the setup where a single laser beam with wavelength $\lambda$, propagating along the positive $z$ axis (see Fig.~\ref{fig1}), is transmitted through a diffusive plate and then focused onto the atomic cloud using a lens with focal length $f$. We assume the lens to be uniformly lit over a circular aperture of diameter $D$.
The simplified numerical procedure employed before for isotropic models~\cite{huntley,modugnomichele} does not apply in this case. The complex amplitude of the speckle field at the position ${\bf r}=(x,y,z)$ measured from the focal point can be computed using the Fresnel diffraction integral~\cite{goodman}:
\begin{multline}
\label{fresnel}
A_1\left(  {\bf r}  \right)  = 
\frac{1}{i\lambda f}  \exp     \left[i 2\pi \left( z + f \right)/\lambda  \right]   \times \\
 \int \int      a_1\left( \alpha, \beta  \right)
\exp  \left[  i\pi \frac{\left( x- \alpha\right)^2 +\left( y- \beta\right)^2 }{\lambda \left( z+f\right)} \right]
\textrm{d}\alpha \textrm{d}\beta,
\end{multline}
where $a_1(\alpha,\beta)$ is the complex field amplitude at the point ${\bf l} \equiv (\alpha,\beta)$ just behind the focusing lens. The potential intensity is $V_1({\bf r}) = \left|A_1\left(  {\bf r}  \right)\right|^2$. Equation~\ref{fresnel} was derived assuming paraxial approximation. Consistently, we will consider only small (positive) displacements from the focal point: $x,y,z \ll f,D$.
A convenient procedure to evaluate the Fresnel integral is to simulate the effect of a large number $N$ of scattering centers randomly placed on the aperture~\cite{goodman,dainty}. On the lens plane, one has: $a_1\left(\alpha, \beta \right) = \sum_{n=1}^{N} o_n \exp(i\phi_n) \delta^{(2)}  \left( {\bf l} -{\bf l}_n \right)$, where ${\bf l}_n \equiv (\alpha_n,\beta_n)$ is the position of the $n$th scatterer, $o_n$ is the modulus of the corresponding scattered wave, and $\phi_n$ is its phase, which has to be sampled from a uniform random distribution in the interval from $-\pi$ to $\pi$. 
To simulate the effect of a uniform illumination, which is the case considered in this Rapid Communication, the moduli $o_n$ have to be identically and independently distributed random variables, while the random positions of the scattering centers must fill the aperture circle uniformly~\cite{SM}.
Substituting the expression for  $a_1\left(\alpha, \beta \right) $ in eq.~\ref{fresnel}, one obtains the complex field $A_1\left(  {\bf r}  \right) $ as the sum of wavelets propagating from the scattering center to the observation point. The field $A_1\left(  {\bf r}  \right)$ then has to be normalized to have the desired average intensity $V_0$.
We verified that for $N\gg100$ the resulting potential has the statistical properties of fully developed speckle patterns~\cite{goodman}. The intensities have the exponential probability distribution $P_{\textrm{bd}}(V)$ defined above.
The spatial correlation function is anisotropic, with cylindrical symmetry around the propagation axis $z$. If one takes two points aligned in the radial direction, the correlation function reads $\Gamma^{\textrm{rad}}(r) = \left[ 2J_1\left(r/\sigma\right) /\left(r/\sigma\right) \right]^2$~\cite{goodman}, where the correlation length is fixed by the parameters of the optical apparatus: $\sigma=\lambda f / \left(D\pi\right)$. 
Instead, in the axial direction the correlation function is~\cite{goodman}: $\Gamma^{\textrm{z}}(r) = \left[\sin(r/\sigma_z)/(r/\sigma_z)\right]^2$, where the axial correlation length is $\sigma_z = 8\lambda (f/D)^2/\pi$. In current experimental implementations, the optical parameters are typically such that $\sigma_z > \sigma$, meaning that the speckle grains are elongated along the beam propagation axis. For example, in the experiment of Ref.~\cite{demarco1} the anisotropy parameter was $\sigma_z/\sigma \approx 6$, while in the later experiment~\cite{demarco2} it was varied in the range $1\lesssim \sigma_z/\sigma \lesssim 10$ by adjusting the aperture of the focusing length~\cite{noteanisotropy}.\\
%
%
%
\begin{figure}
\begin{center}
\includegraphics[width=1.0\columnwidth]{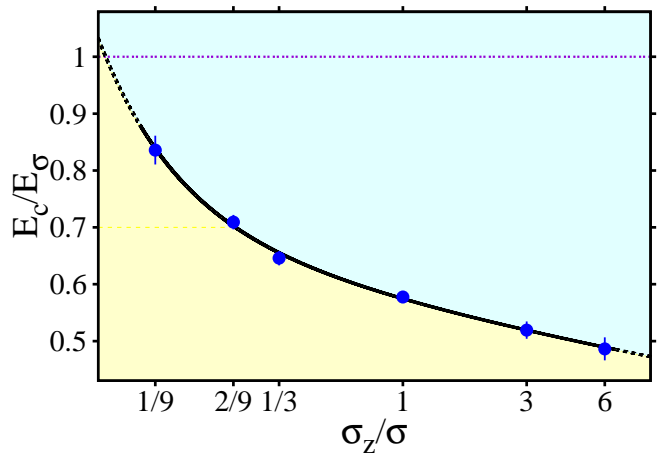}
\caption{(color online) Mobility edge $E_c/E_{\sigma}$ as a function of the anisotropy parameter $\sigma_z/\sigma$. $\sigma_z$ and $\sigma$ are the axial and radial correlation lengths, respectively.
The horizontal purple line indicates the disorder intensity $V_0 = E_\sigma$. The solid black curve is a guide to the eye (the dashed parts are an extrapolation).
}
\label{fig4}
\end{center}
\end{figure}
%
In order to investigate the effects due to the correlation anisotropy, we compute $E_c$ for varying values of the axial correlation length $\sigma_z$, considering both squeezed speckle grains ($\sigma_z/\sigma < 1$) and elongated speckle grains ($\sigma_z/\sigma > 1$). In our computations the box shape is adapted to the disorder anisotropy, see~\cite{SM}.
The disorder intensity is set at $V_0 = E_\sigma = \hbar^2/m\sigma^2$, defined using the (fixed) radial correlation length $\sigma$. 
We find that $E_c$ monotonously decreases as we increase the anisotropy parameter $\sigma_z/\sigma $. In the quasi-isotropic case $\sigma_z/\sigma = 1$, the result agrees with the idealized isotropic model considered above, while it is approximately $50\%$ larger for $\sigma_z/\sigma = 1/9$, and $15\%$ lower for $\sigma_z/\sigma = 6$ (see Fig.~\ref{fig4}).
It is worth noticing that this dependence of $E_c$ on the anisotropy parameter is not trivially related to the scaling of the average correlation energy $E_{{\tilde \sigma}} = \hbar^2 / m {\tilde \sigma}^2$, defined from the geometric mean of the correlation lengths in the three spatial directions: ${\tilde \sigma} =\left( \sigma\sigma\sigma_z\right)^{1/3}$. This suggests that the geometric mean  ${\tilde \sigma}$ is not the unique relevant length-scale, and that the structure of the spatial correlations plays a central role.
We emphasize that in this Rapid Communication we are considering the speckle pattern created by a uniform aperture function. With different kinds of illumination (e.g, the Gaussian illumination~\cite{demarco1,aspect2}), $E_c$ might be somewhat different.\\
While the reduction of $E_c$ due to a large axial correlation length could be observed using currently available experimental setups, the increase of $E_c$ is not easily accessible since the optical apparatuses do not permit us to create squeezed speckle grains.
However, we can show that a similar increase in $E_c$ is induced when two orthogonal speckle patterns are superimposed.
Explicitly, we numerically construct the potential due to the sum of two speckle patterns generated by laser beams with the same wavelength $\lambda$. The first pattern propagates along $z$ and the second along $y$ (see Fig.~\ref{fig1}), and they interfere coherently, as is the case when the two laser beams have the same linear polarization. The total complex amplitude is then~\cite{kirchner,okamoto}: $A_{\textrm{tot}}   \left(   {\bf r}   \right) = 
A_{\textrm{1}}   \left(  {\bf r}  \right)+ A_{\textrm{2}}\left(  {\bf r}  \right).$
The complex amplitude of the second speckle pattern $A_2  \left( {\bf r} \right)$ can be computed using eq.~\ref{fresnel} as described above, just exchanging the roles of the coordinates $y$ and $z$ in the right-hand side. 
For simplicity, we consider two patterns created with equal circular apertures, lit with the same (uniform) intensity, and focused using identical lenses. Thus the corresponding potentials $\left|A_1({\bf r})\right|^2$ and $\left|A_2({\bf r})\right|^2$ have the same radial correlations lengths, which we set at $\sigma \cong 0.75\lambda$. Their axial correlation lengths are extremely large, so that their variations along the respective propagation axes are irrelevant.
This configuration is inspired by the experimental setup of Ref.~\cite{semeghini}.
The potential $V({\bf r}) = \left|A({\bf r})\right|^2$ corresponding to the coherent sum of two blue-detuned fields has the same exponential intensity distribution $P_{\textrm{bd}}(V)$ as a single (blue-detuned) speckle pattern~\cite{goodman}.
The structure of the spatial correlations of this total potential is instead much more intricate~\cite{semeghini}.
To describe it, it is convenient to consider the principal axes $y'$ and $z'$, obtained with a	 $45^\circ$ rotation of the $y$ and $z$ axes around the $x$ axis (see Fig.~\ref{fig1}).
The correlation between two points aligned in parallel with the first principal axis $z'$ is $\Gamma^{z'}(r)=\Gamma^{\textrm{rad}}(r/\sqrt{2})$, meaning that the correlation length is $\sigma_{\textrm{p}} = \sqrt{2}\sigma$~\cite{semeghini}.
Moving in parallel with the second principal axis $y'$, the potential is seen to oscillate rapidly due to the interference fringes and the corresponding correlation function is:
$\Gamma^{y'} (r) = \left[   2J_1\left(  r/  \sigma_{\textrm{p}}  \right) /
\left(  r /  \sigma_{\textrm{p}}  \right)  \cos  \left(  \sqrt{2} \pi r/ \lambda \right)   \right]^2$.
The correlation function along the transverse axis $x$ is instead the same as for a single speckle pattern: $\Gamma^{x}(r)=\Gamma^{\textrm{rad}}(r)$.
For our choice of parameters, the correlation function along the second principal axis $\Gamma^{y'} (r)$ touches zero four times before the first zero of the corresponding function along the transverse axis $\Gamma^{x} (r)$, indicating the strong anisotropy of the disorder correlations.
We consider again a potential with average intensity $V_0 = E_\sigma$.  The corresponding mobility edge is found to be $E_c = 0.67(1)E_\sigma$, significantly higher than for the idealized isotropic disorder. This result is comparable with the one for a single speckle-pattern with squeezed axial correlation length $\sigma_z/\sigma\simeq 1/3$ .
We argue that this increase of $E_c$ is induced by the rapid variations of the potential due to the interference fringes, which effectively reduce the spatial correlation length along the second principal axis.
Experimentalists can easily modify the width of the interference fringes, either by changing the angle between the two beams or by using lasers with different wavelengths. Observing the increase of $E_c$ is thus within experimental reach.\\

We now turn the discussion to the comparison with the available experimental data.
$E_c$ was first measured in Ref.~\cite{demarco1} in the single-pattern configuration. The speckle grains were elongated, corresponding approximately to the anisotropy parameter $\sigma_z/\sigma \approx 6$~\cite{noteanisotropy}. In the regime of disorder strengths $V_0 \approx E_\sigma$, the results were in the range $1.5V_0 \lesssim E_c \lesssim 2V_0$. These findings do not agree with our results for strongly elongated speckle grains: $E_c \approx 0.5V_0$.
Most likely, the reason of this discrepancy traces back to the procedure used to extract the values of $E_c$ from the measurement of the fraction of atoms that remain Anderson localized. In this derivation, the spectral function was approximated using the disorder-free value~\cite{piraud2,piraudthesis}. This approximation is not reliable at the disorder strength necessary to observe Anderson localization. 
Notice also that in the experiment of Ref.~\cite{demarco1} a Gaussian pupil function was employed.
More recently, the mobility edge was measured in the configuration with two crossed speckle patterns created with approximately uniform apertures~\cite{semeghini} . 
For disorder intensities comparable to the correlation energy $V_0 \approx E_\sigma$, the mobility edge was found in the regime $E_c \approx V_0$. This result is significantly larger than the predictions for idealized isotropic models of the disorder and, in this sense, is consistent with our findings. However, it also overestimates our prediction for $V_0 = E_\sigma$. 
This discrepancy is probably due to the fact that in the experiment the two interfering speckle patterns are not equivalent because they were created using slightly different apertures and lenses with different focal lengths. Also, the width of the experimental interference fringes is slightly smaller compared that in to our model.
Furthermore, an exact modeling of the experiment of Ref.~\cite{semeghini} would require us to go beyond the paraxial approximation.
All of these details of the experimental setup, once fully characterized, could be easily implemented in our formalism to compute $E_c$.\\

In conclusion, we have studied the Anderson localization of matter waves exposed to optical speckles in the framework of quantum-chaos theory.
We have shown that the structure of the spatial correlation of the disorder determines the position of the mobility edge, and we have described the effects induced by the correlation anisotropy in concrete experimental configurations, thus paving the way to a quantitative comparison between theory and experiment.
This study sets the basis for future investigations of the effects due to interactions on the transport and on the coherence properties of disordered atomic gases~\cite{pilati2} and on the role played by the fractality of the critical wave functions close the mobility edge~\cite{kravtsov2}.\\

The authors acknowledge fruitful discussions with G. Modugno, A. Scardicchio and G. Semeghini. G. Orso is acknowledged for helpful discussions and for providing the data from Ref.~\cite{orso}. We thank M. Atambo and I. Girotto for their help in the use of the parallel linear algebra software in our numerical calculation.

%
%
%
%
%
%

\newpage


\begin{center}\textbf{Supplemental Material for \\
Anderson localization of matter waves in quantum-chaos theory}\end{center}


%
In the following supplemental material, we provide additional technical details about the numerical procedure we employ to determine 
the energy spectrum and the level-spacing statistics of isotropic and anisotropic speckle patterns.\\

The real-space Hamiltonian of a quantum particle moving in a speckle pattern is given by: $\hat{H} = -\frac{\hbar^2}{2m}\Delta + V({\bf r})$, where $\hbar$ is the reduced Planck's constant, $m$ the particle's mass, and $V({\bf r})$ is the external potential at the position ${\bf r}$ corresponding to the intensity of the optical speckle field. 
We consider a box with periodic boundary conditions and linear dimensions $L_x$, $L_y$ and $L_z$, in the three directions $\iota = x, y, z$.\\
It is convenient to represent the Hamiltonian operator in momentum space as a large finite matrix: $ H_{\mathbf{k},\mathbf{k'}}= T_{\mathbf{k},\mathbf{k'}}+V_{\mathbf{k},\mathbf{k'}}$; $ T_{\mathbf{k},\mathbf{k'}}$ represents the kinetic energy operator, and $V_{\mathbf{k},\mathbf{k'}}$ the potential energy operator.
The wavevectors form a discrete three-dimensional grid: $\mathbf{k}=(k_x,k_y,k_z)$, with the three components $k_\iota=\frac{2\pi}{L_\iota} j_\iota$, where  $j_\iota=-N_\iota/2,...,N_\iota/2-1$, and the (even) number $N_\iota$ determines the size of the grid in the $\iota$ direction and, hence, the corresponding maximum wavevector. Therefore, when expanded in the square matrix format, the size of the Hamiltonian matrix $H_{\mathbf{k},\mathbf{k'}}$ is $N_{\mathrm{tot}}\times N_{\mathrm{tot}}$, where $N_{\mathrm{tot}}= N_x N_y N_z$.\\
In this basis the kinetic energy operator is diagonal:  $T_{\mathbf{k},\mathbf{k'}}=-\frac{\hbar^2 k^2}{2m} \delta_{{k_x},{k_x'}}\delta_{{k_y},{k_y'}}\delta_{{k_z},{k_z'}}$, where $\delta_{{k_\iota},{k_\iota'}}$ is the Kronecker delta.
The element $V_{\mathbf{k},\mathbf{k'}}$ of the potential energy matrix can be computed as: $V_{\mathbf{k},\mathbf{k'}}=\tilde{v}_{\mathbf{k'}-\mathbf{k}}$, where $\tilde{v}_{\mathbf{k}}$ is the discrete Fourier transform of the speckle pattern $V({\bf r})$:
\begin{multline}
\label{fourier}
\tilde{v}_{ k_{x} k_{y} k_{z} }    =   N_{\mathrm{tot}}^{-1}     \sum_{r_x} \, \sum_{r_y} \,  \sum_{r_z}    \\
v_{r_xr_yr_z} \, \exp   \left[- i   \left(   k_{x} r_{x}+  k_{y} r_{y}  +  k_{z} r_{z}  \right)\right]  ;
\end{multline}
here, $v_{{r_xr_yr_z}} = V({\mathbf r=(r_x,r_y,r_z)})$ is  the value of the external potential on the $N_{\mathrm{tot}}$ nodes of a regular lattice defined by $r_\iota=L_\iota n_\iota/N_\iota$, with $n_\iota=0,..,N_\iota-1$. Wavevectors differences are computed exploiting periodicity in wavevector space.\\
%
\begin{figure}
\begin{center}
\includegraphics[width=1.0\columnwidth]{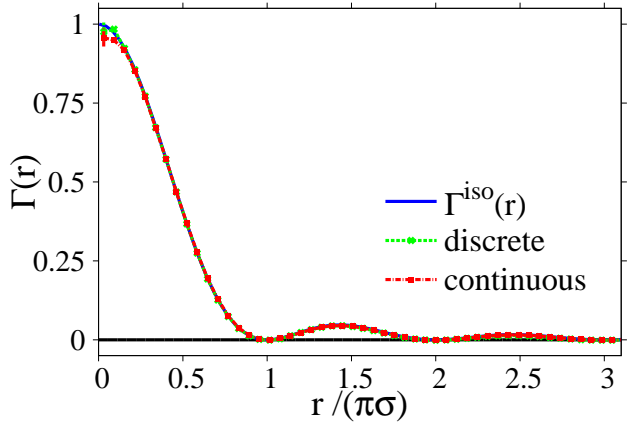}
\caption{Two-point spatial correlation functions of isotropic speckle patterns generated by scattering centers placed on a continuous spherical shell and on a discrete grid.
 The continuous blue curve represents the analytical formula $\Gamma^{\mathrm{iso}}({\bf r})$. The size of the cubic box is $L = 16\pi \sigma$.
 }
\label{figS1}
\end{center}
\end{figure}
%
The numerical algorithm we employ to generate the speckle patterns is described in the main text. It allows us to create both isotropic and anisotropic speckles. 
In the former case, the scattering centers (see main text) have to be placed on a fictitious spherical shell of diameter $D$, with uniform random distribution. Also,  equation (1) of the main text has to be trivially modified, arriving at:
\begin{multline}
\label{fresnel}
A_1\left(  {\bf r}  \right)  = 
\frac{1}{i\lambda f}  \exp     \left[i 2\pi f /\lambda  \right]   \times \\
 \sum_n      o_n \exp \left( i \phi_n \right)
\exp  \left[  i\pi \frac{\left| {\bf r}- {\bf l}_n \right|^2 }{\lambda f} \right] ,
\end{multline}
where ${\bf  l}_n = (\alpha_n,\beta_n,\gamma_n)$ is the position of the $n$-th scattering center on the (fictitious) spherical aperture. One obtains a speckle pattern with the isotropic correlation function $\Gamma^{\mathrm{iso}}({\bf r})$ (see definition in the main text). 
It is worth mentioning that in order to generate isotropic speckle patterns one could instead employ the conventional numerical recipe described in Refs.~\cite{huntleyS,modugnomicheleS}.\\
For isotropic speckles, we employ cubic boxes with $L= L_x=L_y=L_z$ (with periodic boundary conditions).
The numerical recipe described here (and in the main text) creates a potential which does not necessarily satisfy periodic boundary conditions. In principle, this could introduce distortions in the Fourier transform.
However, a speckle pattern compatible with periodic boundary conditions is obtained if the position of the scattering centers is discretized according to:
$\alpha_n \rightarrow  \mathrm{nint} \left( \alpha_n / \delta \right) \delta$, where  $\mathrm{nint}\left( \cdot \right)$ is the function that returns  the whole integer closest to the argument, $\delta = \lambda f/L$ is the discretization step, and the analogous operation is applied to $\beta_n$ and $\gamma_n$.
In figure~\ref{figS1}, we compare the correlation functions numerically measured in speckle patterns generated with continuous and with discrete scatterers, against the analytical formula $\Gamma^{\mathrm{iso}}({\bf r})$. The perfect agreement indicates that the discretization procedure does not alter the statistical properties of the speckles.\\
%
\begin{figure}
\begin{center}
\includegraphics[width=1.0\columnwidth]{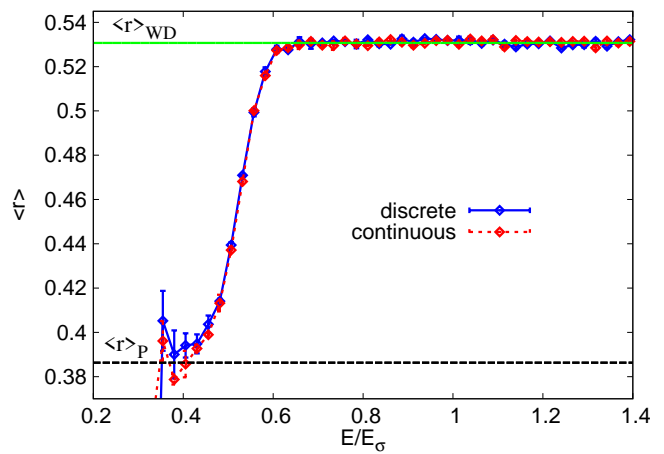}
\caption{ 
Analyses of the level-spacing statistics for isotropic speckle patterns generated by scattering centers placed on a continuous spherical shell and on a discrete grid.
$\left<r\right>$ is the ensemble-averaged adjacent-gap ratio, which is plotted  as a function of the energy $E/E_\sigma$, where $E_\sigma$ is the correlation energy. 
The horizontal green line is the result for the Wigner-Dyson distribution $\left<r\right>_{\textrm{WD}}$, the dashed black line the one for the Poisson distribution  $\left<r\right>_{\textrm{P}}$.
}
\label{figS2}
\end{center}
\end{figure}
%
In figure~\ref{figS2}, we compare the analyses of the energy-levels statistics of two speckle patterns (with average intensity $V_0 = E_\sigma$) generated with continuous and with discrete scatterers. The excellent agreement cross-validates the two procedures, meaning that they are both suitable for our purposes. This is to be expected, since the linear size of the cubic box is $L = 20\pi \sigma$ (this is a typical size we use), much larger than the correlation length of the disorder, so that border effects play a minor role. Notice that  the parameter $\sigma$ characterizes the lengths scale of the disorder spatial correlation, and hence, the typical speckle size. More quantitatively, the full width at half maximum $\ell_c$, defined by the condition $\Gamma^{\mathrm{iso}}(\ell_c/2) =  \Gamma^{\mathrm{iso}}(0)/2$, is  $\ell_c \cong 0.89 \pi \sigma$. The discretization procedure can be easily adapted to have periodicity in anisotropic speckle patterns, provided the focal length $f$ is much larger than the box size.\\
%
\begin{figure}
\begin{center}
\includegraphics[width=1.0\columnwidth]{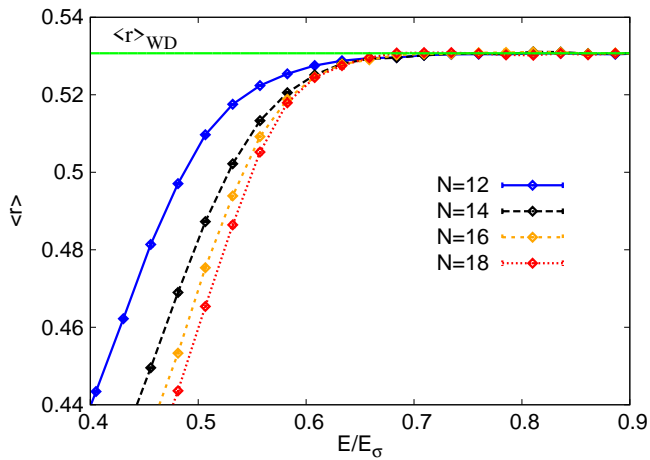}
\caption{
Adjacent-gap ratio $\left<r\right>$ as a function of the energy $E/E_\sigma$, for an isotropic speckle pattern with intensity $V_0 = E_\sigma$. 
The different datasets correspond to different number of wavevectors $N$, thus to different levels of accuracy.
}
\label{figS3}
\end{center}
\end{figure}
%
\begin{figure}
\begin{center}
\includegraphics[width=1.0\columnwidth]{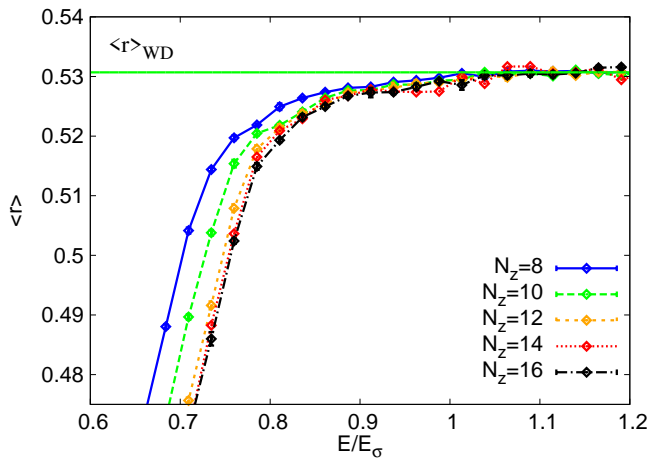}
\caption{Adjacent-gap ratio $\left<r\right>$ as a function of the energy $E/E_\sigma$, for an anisotropic speckle pattern in the single-beam configuration.
The anisotropy parameter is $\sigma_z/\sigma = 2/9$, while the disorder intensity is $V_0 = E_\sigma$. The box sizes are $L_x=L_y=7.5\pi\sigma$ and $L_z = L_x/3$, 
meaning that the box shape has been only partially adapted to the disorder anisotropy.
The different datasets correspond to different number of wavevectors $N_z$ in the axial direction $z$, while in the radial directions the wave-vector number is fixed at $N_x=N_y=12$.
}
\label{figS4}
\end{center}
\end{figure}

A crucial step to guarantee the accuracy of our result is to test that the number of wavevectors (and, correspondingly, the maximum wavevector) is sufficient to have an accurate representation of the speckle pattern and of the orbitals. In figure~\ref{figS3}, we compare the level-spacing statistics for an isotropic speckle pattern obtained with different numbers of wavevectors. For isotropic speckles, it is convenient to set $N_x=N_y=N_z=N$. It is evident that we obtain convergence already for moderately large $N\approx 18$. Consequently, we can afford to perform ensemble averages over a large number of disorder realizations in a suitable computational time. For example, the data corresponding to $N=18$ in figure~\ref{figS3} have been obtained by averaging approximately $3\times 10^4$ disorder realizations, requiring $72$ hours on a 20-cores CPU, using the PLASMA library for linear algebra computations~\cite{plasma}. The maximum wavevector number we consider is $N=40$, allowing us to solve approximately $13$ disorder realizations in 24 hours.
Notice that, when we increase the system size $L$, we proportionally increase $N$, so that the maximum wavevector in the Hamiltonian matrix remains fixed and, therefore, we maintain the same level of accuracy.\\
In the case of anisotropic speckle patterns, it is convenient to (partially) adapt the shape of the box to the disorder anisotropy, so that the system size can be set to be much larger than the disorder correlation length in each direction, without exceedingly increasing the matrix size. Also, the numbers of wavevectors in the three spatial directions $N_x$, $N_y$, and $N_z$ have to be adjusted according to the corresponding linear system sizes, $L_x$, $L_y$, and $L_z$, respectively, and also according to the correlation length in the corresponding direction. The shorter the correlation length, the larger $N_\iota$ is required.
%
\begin{figure}
\begin{center}
\includegraphics[width=1.0\columnwidth]{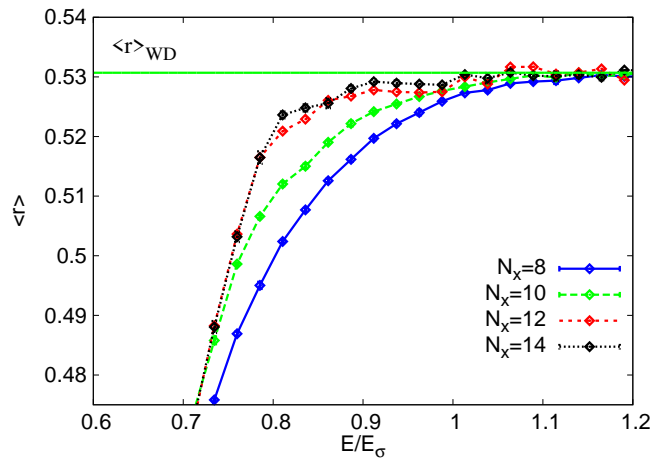}
\caption{
Adjacent-gap ratio $\left<r\right>$ as a function of the energy $E/E_\sigma$, for the same anisotropic speckle pattern as in figure~\ref{figS4}.
Here, the different datasets correspond to different number of wavevectors $N_x=N_y$ in the radial directions $x$ and $y$, while in the axial direction $z$ the wave-vector number is fixed at $N_z = 14$.
}
\label{figS5}
\end{center}
\end{figure}
%
In the single laser-beam configuration, the disorder correlations have axial symmetry around the beam propagation axis $z$; therefore we set $L_x=L_y$ and $N_x=N_y$. 
As an illustrative example, we consider here an anisotropic pattern with strongly squeezed grains corresponding to $\sigma_z/\sigma = 2/9$. The parameter $\sigma_z$ characterizes the correlation length in the axial direction; the full width at half maximum of the axial corresponding correlation function $\Gamma^{z}(z)$ (see definition in the main text) is $\ell_c \cong 0.89 \pi \sigma_z$. For the radial correlation function $\Gamma^{\mathrm{rad}}(r)$, one has $\ell_c \cong 1.029 \pi \sigma$. 
Notice that even in the case $\sigma = \sigma_z$, the disorder is not perfectly isotropic.
In our anisotropic example, we employ a box with an anisotropy which is similar to that of the disorder: $L_z = L_x/3$. It is important to stress that the analysis of the effect due to the finite wavevectors number has to be performed both for the axial  direction $z$ and the radial directions $x$ and $y$, separately. The former effect is analyzed in figure~\ref{figS4}, the latter in figure~\ref{figS5}. Again, we observe convergence with moderately large grid sizes.\\
We emphasize that the analysis of the level-spacing statistic can be used to determine the mobility edge $E_c$ both in the case of isotropic and anisotropic models; see, e.g., Ref.~\cite{schreiberS}. In particular, it was found in Ref.~\cite{schweitzerS} that the position of the mobility edge does not depend on the shape of the box, and that the level-spacing statistics converges in the thermodynamic limit  to the Wigner-Dyson and to the Poisson distributions, in the delocalized and in the localized regimes, respectively. Instead, the level-spacing statistics exactly at the critical point $E = E_c$ (which is system-size independent) was found to depend on the box shape. This does not invalidate our finite-size scaling procedure, but likely implies that the amplitude of the parameter $\left< r \right >$ at $E_c$ for different speckle anisotropies is not universal.
Results analogous to those of Ref.~\cite{schweitzerS} were obtained in Refs.~\cite{braunS,schweitzerS}, where the effect due to the choice of boundary conditions (e.g., periodic boundary conditions vs. hard-wall or Dirichlet boundary conditions) was analyzed. Again, it was found that the position of mobility edge is not affected by the choice of boundary conditions, while the level-spacing distribution at the critical point is.
%
%
\begin{figure}
\begin{center}
\includegraphics[width=1.0\columnwidth]{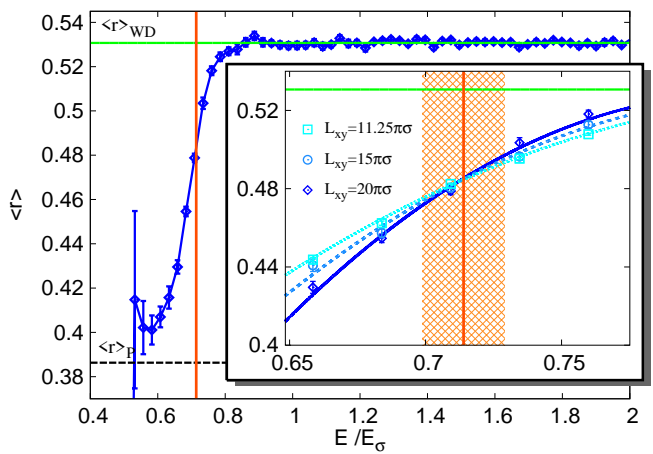}
\caption{
Main panel: ensemble-averaged adjacent-gap ratio $\left<r\right>$ as a function of the energy $E/E_\sigma$ for an anisotropic speckle pattern of intensity $V_0=E_\sigma$ and anisotropy parameter $\sigma_z/\sigma=2/9$.
The different datasets correspond to different box sizes $L_x=L_y$, with $L_z/L_x = 1/3$.
Inset: comparison between different system sizes. The vertical orange line indicates the position of the mobility edge $E_c$ (the rectangle with pattern represents the error-bar). The continuous curves represent the scaling Ansatz $g[x]$ (defined in the main text) expanded to second order.}
\label{figS6}
\end{center}
\end{figure}
%
Figure~\ref{figS6} shows the analysis of the level-spacing statistics for our example of anisotropic speckle, and the finite-size scaling analysis (see inset) employed to locate the mobility edge.
For completeness, we have also analyzed the potential effect on $E_c$ due to the box shape by repeating calculations with different $L/L_z$ ratios. Consistently with the findings of Refs.~\cite{schweitzerS,braunS,schweitzerS}, we also find that there is no systematic bias by changing the box shape; however, larger errorbars are obtained if the shape is not the optimal one, since one has to employ larger matrix sizes.\\
In the configuration with two superimposed speckle patterns, the disorder has the intricate anisotropic spatial correlation structure described in the main text; see figure (1) [panel (c)] and figure (2) in the main text. In this case, we use a cubic box with $L_x=L_y=L_z$, but we rotate the speckle pattern such that  second principal axis $y'$ (see panel (d) of figure (1) in the main text) is aligned with one of the sides of the box, e.g., along the $z$ direction. Then, we increase the wavevector number corresponding to this spatial direction, since the disorder has rapid oscillations and a shorter effective correlation length due to the interference fringes. For the two-pattern configuration described in the main text, we find it is necessary to use a wavevector number $N_z \approx 3N_x$.\\

Finally, it is worth comparing the efficiency of our procedure to determine $E_c$, which is based on the analysis of the level-spacing statistics within quantum-chaos theory, with the one of the transfer-matrix theory employed in Ref.~\cite{orsoS} in the case of isotropic speckle patterns. In the main text we show the quantitative comparison between the two theories both for blue-detuned and red-detuned isotropic speckle fields. The results of transfer-matrix theory have somewhat smaller errorbars, in particular for red-detuned speckles, since in our formalism they require a larger wavevector number, and hence, allow us to consider fewer disorder realizations.
However, we have shown that quantum-chaos theory provides us with effective and flexible tools to address disorder patterns with intricate correlation structures and to adapt the shape of the sample to the disorder anisotropy. More importantly, our exact diagonalization study allows us to disclose interesting properties of the energy spectrum of optical speckles, thus creating a strong link between random matrix theory and ultracold atomic gases.


\begin{thebibliography}{99}
\bibitem{noteanisotropy} The experiments of Ref.~\cite{demarco1,demarco2} implemented a Gaussian illumination, so the identification with the anisotropy parameter used here is only semi-quantitative.
\end{thebibliography}
\end{document}